\documentclass[%
 reprint,
 amsmath,amssymb,
 aps,
 pre,
]{revtex4-2}

\usepackage{graphicx}
\usepackage{dcolumn}
\usepackage{bm}

\begin{document}

\title{Finite curved creases in infinite isometric sheets}

\author{Aaron J. Mowitz}
\email{amowitz@uchicago.edu}
\affiliation{Department of Physics and James Franck Institute, University of Chicago, Chicago, Illinois 60637, USA}

\date{\today}

\begin{abstract}

Geometric stress focusing, e.g. in a crumpled sheet, creates point-like vertices that terminate in a characteristic local crescent shape. The observed scaling of the size of this crescent is an open question in the stress focusing of elastic thin sheets. According to experiments and simulations, this size depends on the outer dimension of the sheet, but intuition and rudimentary energy balance indicate it should only depend on the sheet thickness. We address this discrepancy by modeling the observed crescent with a more geometric approach, where we treat the crescent as a curved crease in an isometric sheet. Although curved creases have already been studied extensively, the crescent in a crumpled sheet has its own unique features: the material crescent terminates within the material, and the material extent is indefinitely larger than the extent of the crescent. These features together with the general constraints of isometry lead to constraints linking the surface profile to the crease-line geometry. We construct several examples obeying these constraints, showing finite curved creases are fully realizable. This approach has some particular advantages over previous analyses, as we are able to describe the entire material without having to exclude the region around the sharp crescent. Finally, we deduce testable relations between the crease and the surrounding sheet, and discuss some of the implications of our approach with regards to the scaling of the crescent size.

\end{abstract}

\maketitle


\section{Introduction}
\label{sec:intro}

Emergent structures are omnipresent in nature. Two-dimensional materials serve as a useful and accessible means to study such structures. Examples include buckling patterns in twisted ribbons under tension \cite{Ribbons}, azimuthal wrinkling in disks floating on a liquid drop \cite{Wrinkling}, and paper origami which can lead to mechanical metamaterials \cite{Origami}. Some emergent structures show the additional property of focusing.  Here deformation is concentrated to indefinitely small regions of the material. While this phenomenon is very commonplace, it goes against our intuition from thermodynamics. If one injects energy into a system, one expects the energy to distribute itself uniformly across the system's size. A paradigmatic example of this localized behavior is seen in fluid boundary layers in pipe flow, where a large amount of viscous drag occurs near the pipe wall \cite{Prandtl}. Another example are the caustic networks of light refracted through a disturbed fluid surface \cite{Caustic}. The size of and the amount of energy stored in these localized regions often exhibit power-law behaviors in a regime where a system parameter is asymptotically small.

One particular phenomenon of interest in this subclass is thin-sheet crumpling. If one compresses a thin sheet, such as a piece of paper, along its boundaries, the sheet begins to buckle. However, instead of exhibiting large wavelength deformations, the material develops highly connected networks of vertices and ridges spanning its entire extent, as seen in the crumpled sheet in Fig. \ref{fig:crescentexamples}. These networks form before the yield stress threshold is reached, but eventually the material plastically deforms, leading to the pattern of creases one sees after unfurling a crumpled piece of paper \cite{BenAmar,WittenRev,GottesmanCommPhys}. Understanding the nature of this stress focusing in thin sheets may provide some much needed insight on boundary layers in elastic materials. Furthermore, being able to manipulate these networks may prove useful for designing localized structures in materials via particular ways of external forcing.

In this paper, we propose a new geometric approach of understanding the extent of stress focusing near a vertex in a crumpled sheet. In Section \ref{sec:dcone}, we review how stress focusing arises in crumpled sheets. We introduce the prototypical model of a vertex and review previous work on determining the size of the stress focused region around a vertex. In Section \ref{curvedcreases}, we introduce our approach. In the prototypical vertex model, one observes a 1-dimensional structure in the shape of a crescent. We argue that this crescent can be described by a separately studied structure known as a curved crease. We outline the geometric constraints which make a curved crease compatible with thin-sheet elasticity. In Section \ref{finitecreases}, we restrict ourselves to creases which more closely resemble the observed crescent. The defining characteristic of this crescent is that it terminates within the material. Furthermore, the extent of the surrounding material is arbitrarily larger than the size of the material. We derive the geometric constraints which are necessary for a curved crease to exhibit these characteristics. In Section \ref{examples}, we construct several examples of curved creases which obey these constraints. In Section \ref{energetics}, we discuss the elastic energy stored in these finite creases, and argue how our approach can avoid the pitfalls of previous approaches. Finally, in Section \ref{discussion}, we discuss predictions of our approach that are experimentally testable. We do not resolve the open question of the extent of stress focusing in a crumpling vertex, but do describe future directions and possible limitations our approach has with regard to the matter.

\section{Stress Focusing}
\label{sec:dcone}

\subsection{Crumpling of thin sheets}
\label{crumpling}

When modeling an initially flat thin elastic sheet with thickness $h$, one can separate its possible deformations into two types: in-plane stretching of the sheet, which leads to strain within the midplane, and out-of-plane bending, which gives rise to curvature of the midplane surface. For a homogeneous isotropic sheet with a Young's modulus $Y$ and Poisson ratio $\nu$, the energy cost due to stretching $E_S$ is given by
\begin{equation}
    E_S = \frac{hY}{2(1-\nu^2)}\int dA\left[\nu(\mathrm{tr}\,\gamma_{ij})^2+(1-\nu)\mathrm{tr}\,\gamma_{ij}^2\right]
\end{equation}
where $\gamma_{ij}$ is the two-dimensional strain tensor of the midplane \cite{AudolyElasGeom}. Similarly, the bending energy $E_B$ is
\begin{equation}
    \label{eq:bendingenergy}
    E_B = \frac{h^3 Y}{24(1-\nu^2)}\int H^2\,dA
\end{equation}
where $H$ is the mean curvature of the midplane surface, defined as the average of the principal surface curvatures \cite{AudolyElasGeom}. There is also a term proportional to the integrated Gaussian curvature $K$, but due to the Gauss-Bonnet theorem, this term does not change with any smooth deformation which does not change the boundary conditions, and so we need not include it \cite{doCarmo}.

In the asymptotic limit where the sheet thickness goes to zero, the elastic energy cost for stretching is prohibitively more expensive relative to bending. This leads to the well-known fact in thin sheet mechanics that a zero-thickness sheet cannot stretch, and are hence commonly referred to as isometric sheets. A geometric consequence of this fact is that the initially flat isometric sheet will have zero Gaussian curvature everywhere after it is deformed. Furthermore, since the Gaussian curvature is the product of the principal surface curvatures, there must be a principal direction of zero curvature at each point in the material. These directions are known as generators, and a surface described by these generators is known as a developable surface \cite{doCarmo}.

Due to the requirement of unstretchability, there are strong constraints on the extent to which one can uniformly confine an isometric sheet. If one tries to uniformly confine the sheet to a region smaller than the size of the sheet $R$, in order to avoid bending of generators, the sheet develops ``vertices" where generators meet at a point within the material, as well as sharp straight creases, or ``ridges," which connect vertices. The sheet is flat everywhere except near these vertices and ridges, and almost all of the elastic stress is localized at these singular structures, due to the large bending deformations in these regions.

A real finite-thickness sheet will also exhibit vertices and ridges, but these regions will be accompanied by some amount of stretching. If the sheet were truly unstretchable, it would cost an infinite amount of bending energy to form vertices and ridges, due to their sharp nature. Allowing for stretching relaxes this divergence. Furthermore, a natural length scale must emerge from the balance between bending and stretching, determining the size of the regions where stress is localized. This has been thoroughly studied for a stretching ridge, where the width has been shown to scale as $h^{1/3}R^{2/3}$ \cite{Lobkovsky}. However, the scaling of stress focusing in a vertex has proven to be much subtler issue, as we shall now see.

\subsection{D-cone core scaling}
\label{dcone}

As with ridges, the size of the region around a vertex where stretching occurs will be finite. There is a straightforward way to study a single vertex in isolation: if one takes a thin sheet and pushes it into a circular ring, one gets a structure known as a developable cone, or d-cone for short \cite{CerdaMahaPRL}. In the zero-thickness limit, the material is unstretched everywhere except at the central forcing point, and has straight generators emanating from this point (hence the name developable). Due to its conical shape and the confining nature of the forcing, the d-cone serves as a paradigm for studying vertices in crumpled sheets, and has been studied extensively both via experiments and simulations \cite{CerdaMahaPRSA,Chaieb,Liang,Wang,Mora,GottesmanNatComm}.

\begin{figure*}
    \includegraphics[width=0.8\textwidth]{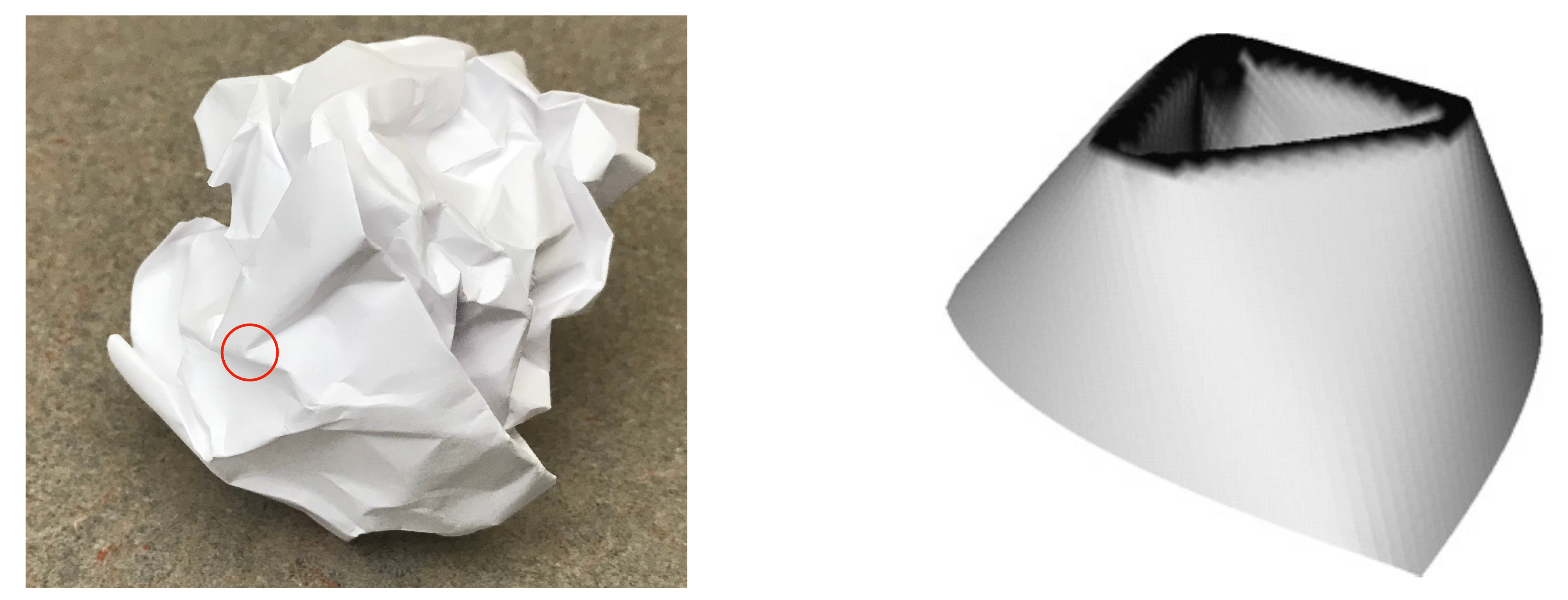}
    \caption{Curved crescents in thin sheets. Left shows a crescent (circled) observed at the vertex of a crumpled sheet. Right shows crescents seen at the corners of a buckled ring ridge (reproduced from \cite{LobkovskyThesis}).}
    \label{fig:crescentexamples}
\end{figure*}

The bending energy of a d-cone can be calculated from the curvature. Due to the locally conical geometry, the curvature goes as $1/r$, where $r$ is the distance from the forcing point, and therefore has a divergence at the center. In order to maintain a finite energy, some distortion from this ideal cone structure is necessary, at least within some small region of size $R_c$. Supposing no appreciable distortion for $r > R_c$, one can then find the scaling of the bending energy:
\begin{equation}
    \label{eq:dconeEB}
    E_B \sim h^3 Y\int_{R_c}^R \frac{1}{r^2}r\,dr \sim h^3 Y\log\frac{R}{R_c}.
\end{equation}
One typically treats $R_c$ as a cutoff radius, separating the outside conical region from the inner region, which is referred to as the d-cone core. In a real finite-thickness sheet, the material will exhibit stretching, which one presumes to be localized to the core region, similar to the vertex in a crumpled sheet. As in the stretching ridge, there will be competition between stretching and bending, leading to stress focusing in the core. This physically manifests itself in the material as a sharp crescent whose length is commonly taken as a measure of $R_c$ \cite{CerdaMahaPRSA}. These crescents are seen in crumpled sheets, as well as at the corners of a buckled ring ridge, as seen in Fig. \ref{fig:crescentexamples}. Because this region is completely localized around the tip, one would expect its size should not depend on any outer dimensions of the sheet, i.e. the size of the sheet $R$ or the size of the confining ring $R_p$. It is therefore expected that $R_c \sim h$, which is supported by a more rigorous energy-balance argument \cite{Liang}.

If one removes a disk of radius $R_c$ from the center of a flat sheet and pushes it into a confining ring, the resulting shape will be that of a d-cone with its tip removed, whose energy cost is given by Eq. \eqref{eq:dconeEB}. This energy alone prefers a large $R_c$, since it will be smaller for larger $R_c$, and its derivative with respect to $R_c$ is independent of $R$. In order for there to an optimal finite $R_c$, there must be an additional contribution to the total energy that prefers small $R_c$. If one now re-attaches the disk to the inner boundary of the truncated d-cone, the disk must deform in order to close the surface and will therefore cost elastic energy. The distortion of the disk may, for example, be similar to that of spherical cap, and so will exhibit a mean curvature of order $1/R_c$. This means that the energy cost due to bending will be of order $h^3 Y(1/R_c)^2 R_c^2\sim R_c^0$, i.e. it will be independent of $R_c$. The distorted disk will also exhibit Gaussian curvature of order $1/R_c^2$, and so will have a stretching energy cost of order $h Y R_c^2$. This energy prefers small $R_c$, and so competes with the energy of the outer region. However, it is independent of $R$, so that if one minimizes the sum of the bending energy of the outer region and the stretching energy of the inner disk, the result is $R_c \sim h$.

The above argument indicates that the core radius should be independent of the outer dimension of the sheet. However, both simulations and experiments have measured an $R_c$ which does depend on the sheet size, and appears to vary as $h^{1/3}R^{2/3}$, the same as a stretching ridge. Furthermore, Cerda and Mahadevan \cite{CerdaMahaPRSA} have proposed a different energy-balance argument that includes stretching due to having a finite-size core, which lengthens the radial generators. This analysis leads to the observed scaling of $R_c$ that depends on $R$.

While it appears the core size scaling is fully explained by this argument, there are still some inconsistencies. The stretching due to lengthening of generators that Cerda and Mahadevan include in their analysis would normally be present in the d-cone outer region in addition to the core. This stretching would have an unphysically large energy cost and would completely dominate the already present bending energy from Eq. \eqref{eq:dconeEB}. It may be possible that the stretching is completely localized to the core region, but no argument has been put forth to justify this \cite{WittenRev}. Furthermore, the d-cone simulations and experiments are limited by the range of sheet thickness they can probe, and typically only use a single order of magnitude range of $h$ when measuring $R_c$. Thus there is still an explanation to be found for the observed scaling of $R_c$. Still, for the moment it appears that the core size increases indefinitely with $R$ for fixed $h$ and decreases indefinitely with $h$ at fixed $R$.

\begin{figure*}
    \includegraphics[width=0.8\textwidth]{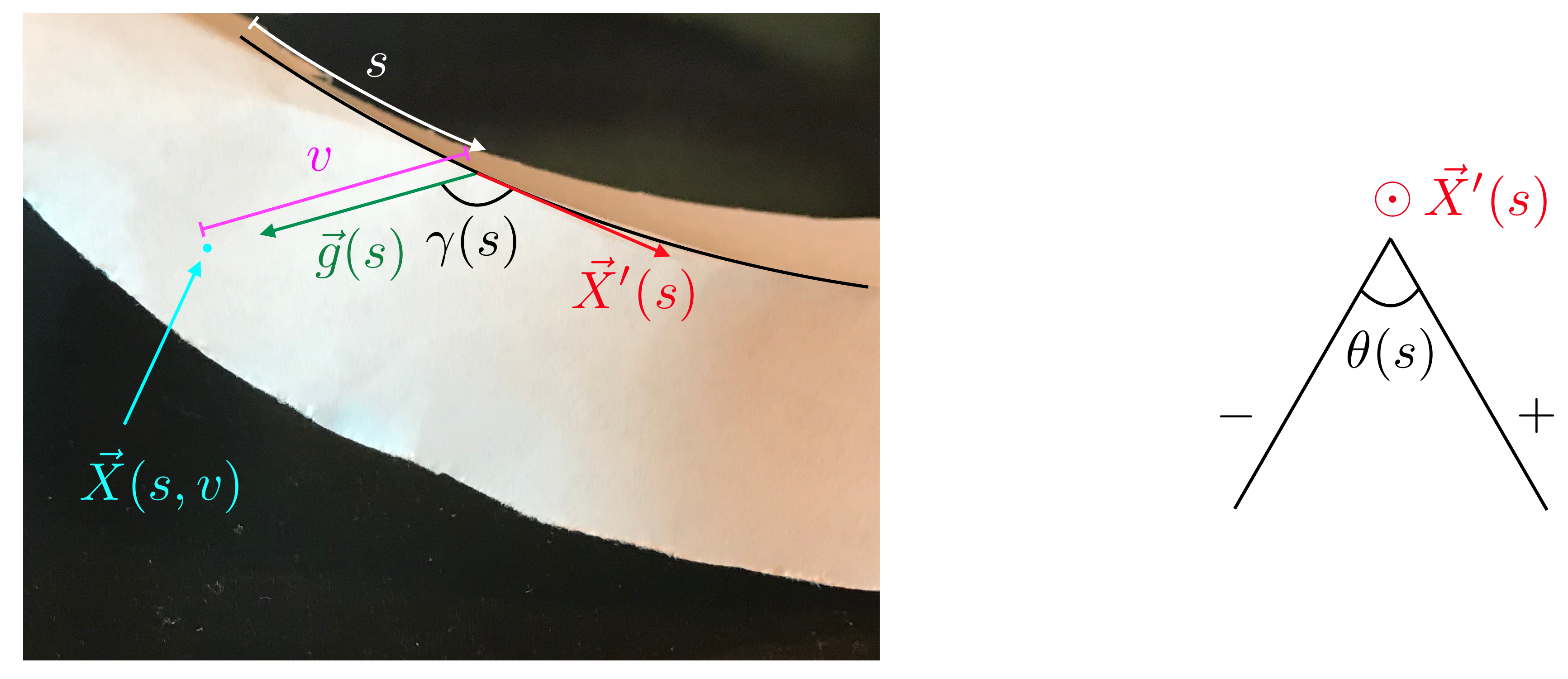}
    \caption{Diagrams labeling quantities pertaining to curved crease. The left shows a view of a narrow curved crease whose crease line is indicated by the black curved line. The right shows a view facing the crease line tangent and indicates the opening angle $\theta$.}
    \label{fig:crease}
\end{figure*}

\section{Curved creases}
\label{curvedcreases}

\subsection{D-cone crescent as a curved crease}
\label{crescentcrease}

We turn back to the observed crescent in the d-cone core. The stretching in the core region occupies a finite area around the crescent. If we take the limit as $h \rightarrow 0$, the crescent will shrink to a point, where the stretching will be localized, just like in a real cone. However, if $R_c$ does have a power law dependence on $R$, we may at the same time take the limit as $R\rightarrow\infty$ in such a way that $R_c$ is kept fixed as $h\rightarrow 0$. In this case, we maintain a crescent of finite size, where the stretching is now localized to the crescent line, so that the material is isometric everywhere else. What we are therefore left with is a sharp curved crease in an isometric sheet. This provides a new way to study the d-cone core region, as we shall see.

Curved creases have been a topic of interest in the field of thin elastic sheets for several decades. If one draws a curved line on a flat piece of paper and folds along that line, the material develops 3-dimensional structure. The curve is no longer planar and the surfaces on either side become curved in a way such that they are compatible with the curvature of the crease line, which is determined by the shape of the planar curve and the opening angle of the fold. These creases have been studied by computer scientists \cite{Huffman} and engineers \cite{DuncanDuncan} as an avenue for designing new geometric shapes. More recently, there has been an interest in curved creases made out of real finite-thickness materials. A simple example is the buckled structure that is formed when one folds along the midline of a circular annulus \cite{DiasPRL,DiasJMPS,DiasEPL}, due to geometric frustration from having a closed curve. The d-cone crescent seems to be a likely candidate as a curved crease, but there are two key differences that separate it from previously studied cases. The crescent is in a sheet of large extent, and the crescent line terminates within the material. There have been studies attempting to model the d-cone that account for either one of these features \cite{Kilian,Farmer,Seffen}, but as of yet, no one has explicitly laid out geometric constraints that yield both of these characteristics.

In order to outline these constraints, we first review what is known about the geometry of curved crease \cite{DuncanDuncan,DiasPRL}. If one draws a curve parametrized by its arc-length $s$ on a flat asymptotically thin ribbon and folds through an opening angle $\theta(s)$ along this curve (see Fig. \ref{fig:crease}), one gets a 3-dimensional structure known as a curved crease. This structure can be thought of as two surfaces meeting a common space curve $\vec{X}(s)$, and since it is formed by an isometric deformation of a flat surface, the two surfaces must be developable. Therefore each surface has a generator at every point, and can be described by the standard parametrization for ruled surfaces:
\begin{equation}
    \label{eq:ruled}
    \vec{X}_\pm(s,v) = \vec{X}(s) + v \hat{g}_\pm(s)
\end{equation}
where $s$ is the arc-length coordinate, $\hat{g}_\pm(s)$ is the direction of a generator at $s$, and $v$ is the coordinate along generators. Here $+$ designates the inner surface inside of the curved crease, and $-$ designates the outer surface. The generators on either side can be described by their angle $\gamma_\pm(s)$ with the crease line, referred to as the generator angle.

Though $\vec{X}_\pm(s,v)$ completely characterizes the crease, one can more easily describe it via the standard language of differential geometry of curves and surfaces. At every point along the crease line, one can define an orthonormal frame $\{\hat{t}(s),\hat{n}(s),\hat{b}(s)\}$, known as the Frenet-Serret frame:
\begin{subequations}
    \label{eq:FrenetFrame}
    \begin{gather}
        \hat{t}(s) \equiv \vec{X}'(s) \label{eq:that}\\
        \hat{n}(s) \equiv \hat{t}'(s)/\|\hat{t}'(s)\| \\
        \hat{b}(s) \equiv \hat{t}(s) \times \hat{n}(s),
    \end{gather}
\end{subequations}
where $\hat{t}$, $\hat{n}$, and $\hat{b}$ are called the tangent, curve normal, and binormal, respectively. This frame allows one to define the intrinsic shape of the curve using two scalar functions: its curvature denoted $\kappa(s)$ and its torsion denoted by $\tau(s)$, which are defined by what are known as the Frenet-Serret formulas \cite{SpivakVol2}:
\begin{subequations}
    \label{eq:Frenet}
    \begin{gather}
        \hat{t}'(s) = \kappa(s) \hat{n}(s) \label{eq:tprimefren}\\
        \hat{n}'(s) = -\kappa(s) \hat{t}(s) + \tau(s) \hat{b}(s) \\
        \hat{b}'(s) = -\tau(s) \hat{n}(s).
    \end{gather}
\end{subequations}

\begin{figure}
    \includegraphics[width=0.4\textwidth]{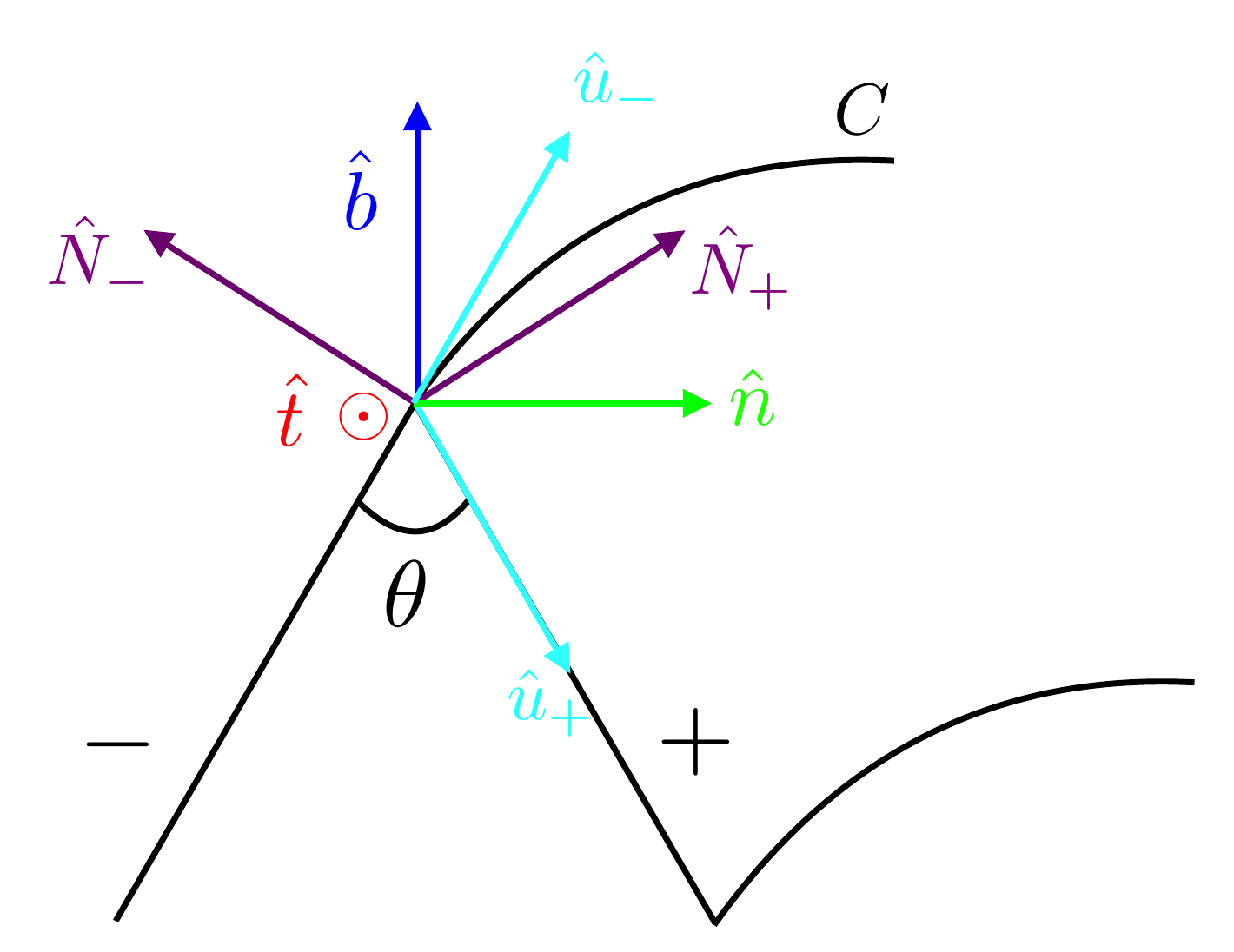}
    \caption{Co-ordinate bases for curved creases. The crease line marked $C$ is directed out of the page at one point, and is bending to the right. The Frenet-Serret basis ($\hat{t}$, $\hat{n}$, $\hat{b}$) of the crease line is shown at this point. The surfaces form an opening angle $\theta$, and are marked $+$ and $-$ with their surface normals $\hat N_\pm$.  The Darboux bases of each side ($\hat{t}$, $\hat{u}_\pm$, $\hat{N}_\pm$) are shown.}
    \label{fig:geodesiccurvature}
\end{figure}

We also need to characterize the geometry of the surfaces on either side of the crease line. Since $\vec{X}(s)$ can be seen as a curve lying in either surface of the crease, one can also define another orthonormal frame, $\{\hat{t}(s),\hat{u}_\pm(s),\hat{N}_\pm(s)\}$, known as the Darboux frame, where $\hat{N}_\pm(s)$ is the surface normal along the crease line, and $\hat{u}_\pm(s)\equiv \hat{N}_\pm(s)\times\hat{t}(s)$ is the tangent normal. Similar to the Frenet-Serret case, one can define three scalar quantities $\kappa_{g\pm}(s)$, $\kappa_{N\pm}(s)$, and $\tau_{g\pm}(s)$ from the rate-of-change of the Darboux frame \cite{SpivakVol3}:
\begin{subequations}
    \label{eq:Darboux}
    \begin{gather}
        \hat{t}'(s) = \kappa_{g\pm}(s) \hat{u}_\pm(s) + \kappa_{N\pm}(s) \hat{N}_\pm(s) \label{eq:tprimedarb}\\
        \hat{u}_\pm'(s) = -\kappa_{g\pm}(s) \hat{t}(s) + \tau_{g\pm}(s) \hat{N}_\pm(s) \label{eq:uprimedarb}\\
        \hat{N}_\pm'(s) = -\kappa_{N\pm}(s) \hat{t}(s) - \tau_{g\pm}(s) \hat{u}_\pm(s). \label{eq:Nprimedarb}
    \end{gather}
\end{subequations}
$\kappa_{g\pm}(s)$ and $\kappa_{N\pm}(s)$ are known as the geodesic and normal curvatures, respectively, and are the components of the curvature of $\vec{X}(s)$ tangent and normal to the surface. $\tau_g(s)$ is known as the geodesic torsion, and measures how much the curve is twisting out of the surface tangent plane. Fig. \ref{fig:geodesiccurvature} shows an example of these frames for a curved crease.

\subsection{Constraints due to isometricity}
\label{isometricity}

Although the above provides a complete geometric description of a curved crease, we have not imposed any constraints due to the crease being formed by an isometric deformation of a flat sheet. Before one folds along the crease line, the crease curve is planar (i.e. has no torsion), and only has geodesic curvature. Once one has made the fold, the crease curve may develop torsion, and will have both normal and geodesic curvature when seen as a curve in either the inner or outer surface. However, the geodesic curvature does not change under isometric deformations, so the geodesic curvature as measured in the inner surface ($\kappa_{g+}$) must equal that measured in the outer surface ($\kappa_{g-}$), and must be the same as the curvature of the originally drawn crease line. This puts a constraint between the geometry of the crease line and the shape of either side of the crease \cite{DuncanDuncan,Fuchs}:
\begin{subequations}
    \label{eq:geod}
    \begin{gather}
        \kappa_g(s) = \kappa(s)\sin\theta(s)/2 \\
        \kappa_{N\pm}(s) = \pm\kappa(s)\cos\theta(s)/2 \\
        \tau_{g\pm}(s) = \tau(s) \pm \theta'(s)/2.
    \end{gather}
\end{subequations}
We refer to this as the ``compatibility constraint." It is worth noting that as a consequence of this constraint, the binormal vector of the curve must bisect the opening angle of the crease, as seen in Fig. \ref{fig:geodesiccurvature}. This means that when one forms a crease with a particular opening angle, the direction of the curvature of the crease curve is determined.

We mentioned earlier that the surfaces on either side must be developable. Even though it is a well-known fact that developable surfaces are ruled \cite{SpivakVol3}, the parametrization given by Eq. \eqref{eq:ruled} does not guarantee developability. In order to do so, the directions in which $\hat{g}_\pm(s)$ point must be constrained by the condition that $\vec{X}'(s)$, $\hat{g}_\pm(s)$, and $\hat{g}_\pm'(s)$ are coplanar \cite{doCarmo}. This leads to a relation between the direction of the generators and the twisting of the crease line \cite{DuncanDuncan,Fuchs}:
\begin{equation}
    \label{eq:develop}
    \tan\gamma_\pm(s) = -\frac{\kappa(s)\cos\theta(s)/2}{\tau(s)\pm\theta'(s)/2}
\end{equation}
which we call the ``developability constraint."

With these constraints in hand, it is helpful to know how many and what quantities need to be specified in order to completely determine the shape of a crease. Intuitively, if one folds along a line with a given curvature profile, and the opening angle is prescribed as a certain profile everywhere along the fold line, it seems the entire shape of crease should be determined. This is true if one also considers the bending energy of the surface, but the geometric constraints alone are not enough. We can see this by counting the number of quantities needed to describe the crease, and comparing it to the number of constraints.

The shape of the crease line $\vec{X}(s)$ is fully specified by its curvature $\kappa(s)$ and torsion $\tau(s)$. The outer and inner surfaces are defined by their generators, whose directions in space are specified by the generator angles $\gamma_\pm(s)$ and the opening angle $\theta(s)$. Finally, one must know the curvature of the crease line within either surface, i.e. the geodesic curvature $\kappa_g(s)$. We therefore have a total of six quantities to be specified. However, this number will be reduced by the isometric constraints. The compatibility constraint will reduce the number by one, as it is a single constraint relating the orientation of the crease line and the opening angle. The developability constraint is a single constraint on each surface that requires them to be developable, and so reduces the number further by two. One therefore needs to specify three quantities of the original six to determine the shape of an isometric crease.

For example, suppose one specifies $\kappa_g$, $\tau$, and $\theta$. Eq. (\ref{eq:geod}a) will then determine $\kappa$, so that the crease line can be found, and Eq. \eqref{eq:develop} will yield the generator angles, so that the generators on both sides will be known. Alternatively, if one includes the condition of minimum energy, then this reduces the number of necessary quantities to two. One can then specify just $\kappa_g$ and $\theta$, which is then consistent with the expectation mentioned above.

As an alternative approach, one could start by specifying the surface on, say, the inner side, assuming it is developable. If one constrains a curve with a particular geodesic curvature profile to lie on this surface, its curvature and torsion will be determined. Furthermore, the generator angle on the inner surface will also be known without reference to the other side of the crease. These three quantities are then sufficient to specify the crease. First, $\theta$ may be found from Eq. (\ref{eq:geod}a). Then Eq. \eqref{eq:develop} can give the generator angle on the outer surface. These two procedures will be useful when constructing particular examples of curved creases in Section \ref{examples}.

As noted in Section \ref{crumpling}, the energy content of isometric surfaces, such as those on either side of a crease, depends on their mean curvature $H(s, v)$, as shown by Eq. \eqref{eq:bendingenergy}. It can be shown that the two ruled surfaces on either side of a crease each have a unique curve $\vec{\sigma}_\pm(s)$, known as the striction curve \cite{doCarmo}, defined by the condition $\vec{\sigma}_\pm'(s)\cdot\hat{g}_\pm'(s)=0$. Geometrically, $\vec{\sigma}_\pm(s)$ is the point along $\hat{g}_\pm(s)$ which is closest to the ruling $\hat{g}_\pm(s+ds)$ in the limit $ds\rightarrow 0$. When the surfaces are developable, as they are in our case, the points on the striction curve are singular points of the surfaces, i.e. $\hat{g}_\pm(s)$ and $\hat{g}_\pm(s+ds)$ intersect at $\vec{\sigma}_\pm(s)$ in the limit $ds\rightarrow 0$. One can then show that the mean curvature of the surface at the coordinates $(s,v)$ will be inversely proportional to $d_\pm(s) + v$, where $d_\pm(s)$ is the distance on the surface from the crease point $s$ to the striction curve $\vec{\sigma}_\pm(s)$. We prove this in Appendix \ref{meancurvscaling}. We note that this assumes the striction curve, and therefore the singular points, does not lie in the material, so that $\vec{\sigma}_\pm(s)=\vec{X}(s) - d_\pm(s)\hat{g}_\pm(s)$. We validate this assumption in Section \ref{addconstraints}.

From the defining condition $\vec{\sigma}_\pm'(s)\cdot\hat{g}_\pm'(s)=0$, one can show \cite{doCarmo}
\begin{equation}
    \label{eq:dpm}
    d_\pm(s)=\frac{\vec{X}'(s)\cdot\hat{g}'_\pm(s)}{\|\hat{g}'_\pm(s)\|^2}.
\end{equation}
We can then explicitly find $d_\pm(s)$ in terms of quantities describing the crease. Using the Darboux frame, we can write $\hat{g}_\pm(s) = \cos\gamma_\pm(s)\vec{X}'(s) \pm \sin\gamma_\pm(s)\hat{u}_\pm(s)$. After take the derivative of $\hat{g}_\pm(s)$ and using Eq. \eqref{eq:Darboux}, we can substitute the result into Eq. \eqref{eq:dpm}, yielding $d_\pm(s)=\mp\sin\gamma_\pm(s)/(\kappa_g(s)\pm\gamma_\pm'(s))$.

In order to find the exact expression for $H(s,v)$, we can calculate the mean curvature at a given $v$, such as the crease line, where $v=0$. The mean curvature then takes the form $H(s,v) = H(s,0)/(1+v/d(s))$. To determine $H(s,0)$, we may use Euler's theorem, which for a chosen direction, expresses the surface normal curvature $\kappa_N$ in terms of the principal curvatures $\kappa_0,\,\kappa_1$ and the angle between our chosen direction and a principal one $\eta$ \cite{SpivakVol3}: $\kappa_N = \kappa_0\cos^2\eta + \kappa_1\sin^2\eta$. If we choose the direction to be the crease line tangent, then $\kappa_N$ is the normal curvature of the crease line, and if we choose the $\kappa_0$ direction to be the generator at $s$, then $\kappa_0$ is zero and $\eta$ is the generator angle $\gamma(s)$. So the mean curvature at the crease line is $H(s,0) = \kappa_1(s)/2 = (\kappa_N(s) \csc^2\gamma(s))/2$. We can then generalize to both surfaces of our crease, and find the mean curvature at a general point in either surface:
\begin{equation}
    \label{eq:meancurv}
    H_\pm(s,v) = \frac{\kappa_{N\pm}(s)\csc\gamma_\pm(s)}{2(\sin\gamma_\pm(s)\mp v(\kappa_g(s)\pm\gamma_\pm'(s)))}.
\end{equation}
We note that in the orientation of Fig. \ref{fig:geodesiccurvature}, $H(s, v)$ is necessarily positive (concave) on the inner surface and negative (convex) on the outer one.

We can now determine the bending energy of a curved crease using Eq. \eqref{eq:bendingenergy}. For our parametrization $\vec{X}_\pm(s,v)$, the area element $dA$ can be written as $\|\partial_s\vec{X}_\pm \times \partial_v\vec{X}_\pm\|\,ds\,dv$ \cite{doCarmo}. Then the amount of bending energy stored in the sector of the sheet from a segment of the crease of length $ds$ is given by
\begin{equation}
    dE = \frac{B}{2}\,ds\,\sum_\pm \int H_\pm^2(s,v)\, \|\partial_s\vec{X}_\pm \times \partial_v\vec{X}_\pm\|\,dv,
\end{equation}
where $B\equiv h^3Y/(12(1-\nu^2))$ is the bending stiffness. Calculating $\|\partial_s\vec{X}_\pm \times \partial_v\vec{X}_\pm\|$ and using our expression for $H_\pm(s,v)$ yields \footnote{We note that this expression has disagreement with the energy reported in \cite{DiasPRL}, which has not been fully resolved}
\begin{equation}
    dE = \frac{B}{2}\,ds\,\sum_\pm \int \frac{(\tau\pm\theta'/2)^2 \sec^2\gamma_\pm}{4(\sin\gamma_\pm \mp v(\kappa_g\pm\gamma_\pm'))}\,dv.
\end{equation}
The integrand has the form $A/(B + v C)$, where $A$, $B$, and $C$ are independent of $v$. Thus the $dv$ integral from 0 to $R$ is $(A/C) \log ((B + RC)/B + R C)$:
\begin{widetext}
\begin{equation}
    \label{eq:creaseenergy}
    dE = \frac{B}{8}\,ds\,\sum_\pm \mp\frac{(\tau\pm\theta'/2)^2 \sec^2\gamma_\pm}{\kappa_g\pm\gamma_\pm'}\log\left(\frac{\sin\gamma_\pm \mp R(\kappa_g\pm\gamma_\pm')}{\sin\gamma_\pm}\right).
\end{equation}
\end{widetext}

\section{Finite curved creases in sheets of infinite extent}
\label{finitecreases}

\subsection{Additional constraints}
\label{addconstraints}

\begin{figure}
    \includegraphics[width=0.48\textwidth]{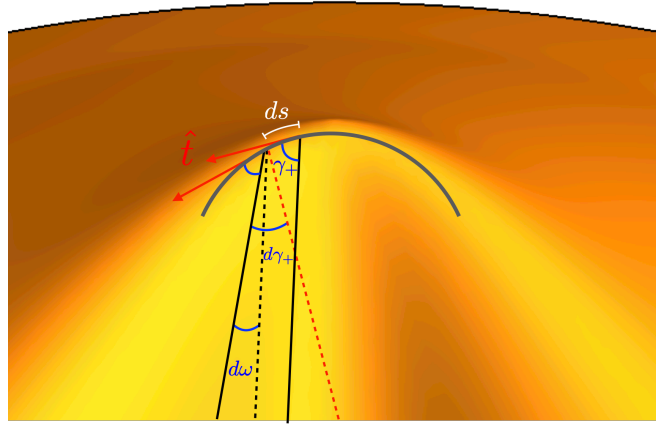}
    \caption{Rotation of an inner surface generator as one moves along the crease line. As one moves along the crease line (gray curved line), a material generator (solid black lines) will rotate in the surface, measured by the angle $d\omega$ between the original generator (dotted black line) and its neighbor. This rotation can be decomposed into two parts. If one transports the original generator along the crease while keeping the generator angle $\gamma_+$ the same, the generator rotates due to the curvature of the crease and aligns with the red (gray) dotted line. This line then rotates in order to line up with the neighboring generator, which comes from the generator angle changing along the crease. The angle of this rotation is therefore $d\gamma_+$.}
    \label{fig:splay}
\end{figure}

The discussion in the previous section presumed that the surface defined by the generators was smooth. Though this is true for a sufficiently narrow region around the crease. It is not in general true for the infinite sheets we consider here. The requirement that this infinite region be isometric imposes additional constraints on the crease itself. Although a set of generators defined along the crease curve may define a smooth surface nearby, this does not preclude the possibility of nearby generators intersecting each other, which would lead to singularities in the sheet. If we consider narrow sheets as we did above, this is not an issue, as we can take our sheet to be narrow enough so that neighboring generators do not have the chance to cross. However, if we consider infinitely large sheets, we must impose a condition that prevents intersecting generators. In order to do this, let us consider two neighboring generators in the inner surface along $\hat{g}_+(s)$ and $\hat{g}_+(s+ds)$, as shown in Fig. \ref{fig:splay}. As one moves along the crease line from $s$ to $s + ds$, $\hat{g}_+(s)$ will rotate in the surface by an angle $d\omega$ to align with $\hat{g}_+(s+ds)$ \footnote{We must emphasize that we are forbidding generators from intersecting within the curved material. It is therefore important to note that since the material is curved, $\hat{g}_+(s)$ and $\hat{g}_+(s+ds)$ lie in different tangent planes of the surface. This means the intrinsic rotation angle $d\omega$ between them is not the space Euclidean angle, i.e. the angle spanned by the intersecting lines in space. Instead, the intrinsic angle is measured as shown in Fig. \ref{fig:splay}, by parallel transporting $\hat{g}_+(s)$ from the point $s$ to the point $s+ds$, and then measuring the angle spanned by this transported generator and $\hat{g}_+(s+ds)$. However, one can measure $d\omega$ much more simply if one draws the generators on the material and then flattens the surface. Since the surface is developable, the intrinsic angle between neighboring drawn-on generators in the curved state will remain unchanged when the surface is flattened. Then one can measure $d\omega$ in the flat state without having to parallel transport, as the intrinsic angle and space Euclidean angle will be the same.}. To avoid the line along $\hat{g}_+(s+ds)$ intersecting with the line along $\hat{g}_+(s)$, the rotation must be clockwise, so that $d\omega<0$. This rotation can be broken down into two components. If we keep the generator angle $\gamma_+(s)$ the same as we rotate $\hat{g}_+$ from $s$ to $s+ds$, the generator rotates due to the curvature of the crease line. This rotation is how much the curve tangent rotates in the material, given by the geodesic curvature $\kappa_g(s)\,ds$. Then if we rotate $\hat{g}_+$ at $s+ds$ to its new generator angle $\gamma_+(s+ds)$, the generator rotates due to the generator angle changing along the curve, given by $d\gamma_+(s) = \gamma_+'(s)\,ds$. The sum of these will be the total rotation of the generator $d\omega$, so we must have $\kappa_g(s)+\gamma_+'(s)<0$. A similar argument can be carried out for generators in the outer surface, leading to $\kappa_g(s) -\gamma_-'(s)>0$. These two inequalities can be written together as 
\begin{equation}
    \label{eq:posplay}
    \gamma_\pm'(s) < \mp\kappa_g(s),
\end{equation}
and gives us a constraint at each point along the crease line that prohibits neighboring generators from intersecting. A similar inequality is presented in \cite{DiasPRL} for creased ribbons, and agrees with ours in the limit as the ribbon width goes to infinity. We will refer to this inequality throughout as the ``positive splay condition."

Up until now we have considered creases in which the crease curve spans the entirety of the material surface. However, it is possible for the crease curve to have a length smaller than the size of the sheet, such as the crescent observed in the core of a d-cone, so that it is localized to a small region, and therefore must terminate within the material. This termination must occur in such a way that the surrounding surface remains smooth and continuous. We will now discuss the conditions that enforce this.

We consider a termination point on the crease at a location denoted by $s_t$, and look at how the previously discussed quantities must behave at this termination point. As one approaches $s_t$, the generators on either side must become more and more parallel so that the two surfaces of the crease meet at a common generator to form a continuous surface. This means that generator angles must be equal and opposite at the termination point: $\gamma_+(s_t) = -\gamma_-(s_t)$. However, these generator angles cannot be negative, as that would lead to the generators crossing from one side of the crease to the other. Therefore, we must have $\gamma_\pm(s_t) = 0$. Geometrically this means that the common generator at $s_t$ must be parallel to the curve tangent. Furthermore, as one crosses this common generator, the surface normal must be continuous. This will be true if as we approach $s_t$, the tangent planes of each side become more parallel and the opening angle becomes larger and larger, so that $\theta(s_t)=\pi$.

We can say more about the behavior of $\tau$ and $\theta'$ at $s_t$ by using the positive splay condition. We first expand our developability constraint, Eq. \eqref{eq:develop}, near $s_t$:
\begin{equation}
    \label{eq:gammanearst}
    \gamma_\pm \rightarrow \frac{\kappa_g\cdot (\theta-\pi)}{2(\tau\pm\theta'/2)}.
\end{equation}
If the leading order behaviors of $\tau$ and $\theta$ near $s_t$ are $\tau(s)\sim A \,(s-s_t)^\alpha$ and $\theta(s) \sim B \,(s-s_t)^{\beta+1} + \pi$ (so that $\theta'(s)\sim(\beta+1) B\, (s-s_t)^\beta$), we see that
\begin{equation}
    \gamma_\pm(s)\sim \frac{\kappa_g(s_t)\cdot B\,(s-s_t)^{\beta+1}}{2(A\,(s-s_t)^\alpha \pm (\beta+1)B\,(s-s_t)^\beta/2)}.
\end{equation}
If we then differentiate and normalize by $\kappa_g(s_t)$, we find
\begin{equation}
    \Gamma_\pm'(s) \sim \frac{(\beta+1)B^2(s-s_t)^{2\beta}\mp2AB(\alpha-\beta-1)(s-s_t)^{\alpha+\beta}}{(2A\,(s-s_t)^\alpha \pm (\beta+1)B(s-s_t)^\beta)^2},
\end{equation}
where $\Gamma_\pm' \equiv \gamma_\pm'/\kappa_g$. We note that the possibility $\beta > \alpha$ is ruled out, as it entails that $\Gamma_\pm'\rightarrow 0$, and thus violates the positive splay condition of Eq. \eqref{eq:posplay} on the inner surface, which requires $\Gamma_+' < -1$ everywhere. Furthermore, when $\beta<\alpha$, $\Gamma_\pm'(s)\sim 1/(\beta+1)$, which also does not satisfy the positive splay condition, since $\beta$ is non-negative. Therefore, in order to avoid intersecting generators near $s_t$, we must have $\beta=\alpha$. In other words, the ratio of $\tau/\theta'$ must approach a constant $C$ as one approaches $s_t$. The value of $C$ will be constrained by the positive splay condition, and will depend on the exact behavior of $\theta$ and $\tau$ near $s_t$, as we will see in Section \ref{conseq}. We can then summarize the conditions necessary for a continuous surface at $s_t$:
\begin{gather}
    \gamma_\pm \rightarrow 0 \label{eq:gammast}\\
    \theta \rightarrow \pi \label{eq:thetast}\\
    \tau / \theta' \rightarrow C \label{eq:tautheta}.
\end{gather}
We note that Eq. \eqref{eq:develop} implies that the first constraint will be automatically satisfied if the second constraint is true, so it suffices to impose Eqs. \eqref{eq:thetast} and \eqref{eq:tautheta}.

It is reasonable, though not necessary, to also require that the surface of the crease is smooth everywhere. This means the mean curvature will be continuous everywhere, particularly at the joining generator. At $s_t$, Eq. \eqref{eq:meancurv} is indeterminate due to our earlier conditions, as $\kappa_{N\pm}(s_t)$ is zero while $\csc\gamma_\pm(s_t)$ is infinite. This indeterminacy can be resolved by expressing $H_\pm$ in terms of $\tau$ and $\theta'$ using Eq. \eqref{eq:develop}. If we then expand the result around $s_t$, we get
\begin{equation}
    \label{eq:Hnearst}
    H_\pm(s_t,v) \rightarrow \mp\frac{\tau(s_t)\pm\theta'(s_t)/2}{2(\gamma_\pm(s_t)\mp v(\kappa_g(s_t)\pm\gamma_\pm'(s_t))}.
\end{equation}
We can impose the continuity condition at $s=s_t$. However, as mentioned at the end of Section \ref{isometricity}, the mean curvature of the outer surface is positive while the mean curvature of the inner surface is negative. Therefore, in order for $H_\pm$ to be continuous along the line at which the two sides meet, the mean curvature must be zero at $s_t$. Assuming $\kappa_g$ is finite at the termination point, this will be true if both $\tau$ and $\theta'$ vanish at $s_t$:
\begin{gather}
    \theta' \rightarrow 0 \label{eq:thetaprimest}\\
    \tau \rightarrow 0. \label{eq:taust}
\end{gather}

\subsection{Consequences}
\label{conseq}

Even under the non-essential assumptions of nonvanishing $\kappa_g$ and continuous mean curvature $H$, the termination conditions of Eqs. (\ref{eq:gammast}-\ref{eq:tautheta}) and (\ref{eq:thetaprimest}-\ref{eq:taust}) impose only weak restrictions on the crease parameters $\theta$ and $\tau$. Here we verify that these termination conditions are generally sufficient to give concrete realizations of surfaces with terminating creases. In order to see some of the consequences of having a terminating crease, let us assume the simplest scalings one can have for $\theta$ and $\tau$ that obey the above conditions, by doing an expansion about the termination point $s_t$ for the generic case where $\kappa_g(s_t) \neq 0$:
\begin{gather}
    \theta(s) = \pi + \Theta''\,(\Delta\tilde s)^2/2 + \mathcal{O}((\Delta \tilde{s})^4) \\
    \tau(s) = C \,\Theta''\, \kappa_g(s_t) \Delta\tilde s + \mathcal{O}((\Delta \tilde{s})^3)
\end{gather}
where $\Theta''\equiv\theta''(s_t)/\kappa_g^2(s_t)$ and $\Delta\tilde s \equiv \kappa_g(s_t)\cdot(s-s_t)$. In order to find $\gamma_\pm(s)$ around $s_t$, we can substitute the above scalings into Eq. \eqref{eq:gammanearst}:
\begin{equation}
    \gamma_\pm(s) \approx \frac{1}{2(2C\pm1)}\Delta\tilde s + \mathcal{O}((\Delta \tilde{s})^3).
\end{equation}

As mentioned in Section \ref{addconstraints}, the values $C$ can take are constrained by the positive splay condition. We can see this by differentiating the above and using the positive splay condition:
\begin{equation}
    \label{eq:nondimsplay}
    \frac{1}{2(2C\pm1)} < \mp 1.
\end{equation}
We first consider the more restrictive requirement on the inner surface, corresponding to the upper signs. In order for this inequality to be satisfied on the inner surface, the left hand side must be negative, so that $C<-1/2$. Furthermore, we can rearrange the inequality to get $C>-3/4$, so that $C$ is bounded both below and above: $-3/4<C<-1/2$. Note that this automatically satisfies Eq. \eqref{eq:nondimsplay} on the outer surface. These bounds will be important when considering specific examples in Sec. \ref{examples}.

\subsection{Finite symmetric crease}
\label{finitesymmcrease}

So far we have considered the behavior of a crease near a termination point in the material. However, in order to know the energy cost needed to form the crease, we must consider its global behavior. Since the d-cone core crescent is symmetric, we will consider symmetric creases which have two termination points. We choose $s=0$ to be the symmetry point, so that the termination points are at $\pm s_t$. In addition to conditions at the termination points, there are also conditions that $\tau,\,\theta'$, and $\gamma_\pm$ must satisfy at the symmetry point. By symmetry, $\tau(s)$ and $\theta'(s)$ must be odd functions, and therefore must be zero at $s=0$. The generator at $s=0$ also has to be perpendicular to the crease, so that $\gamma_\pm(0)=\pi/2$, which is actually automatically true by Eq.  \eqref{eq:develop}. Finally, we must satisfy the positive splay condition. As seen in Section \ref{conseq}, we may expand Eq. \eqref{eq:develop} (this time around $s=0$), differentiate with respect to $s$, and write the result in non-dimensional form:
\begin{equation}
    \Gamma_\pm'(0) \approx \frac{\tau'(0)\pm\theta''(0)/2}{\kappa_g^2(0)\cot \theta_0/2}
\end{equation}
where $\theta_0$ is the opening angle at $s=0$. This must obey $\Gamma_\pm' < \mp 1$.

\section{Examples}
\label{examples}

\subsection{Crease with constant geodesic curvature}
\label{constantkappag}

\begin{figure*}
    \includegraphics[width=\textwidth]{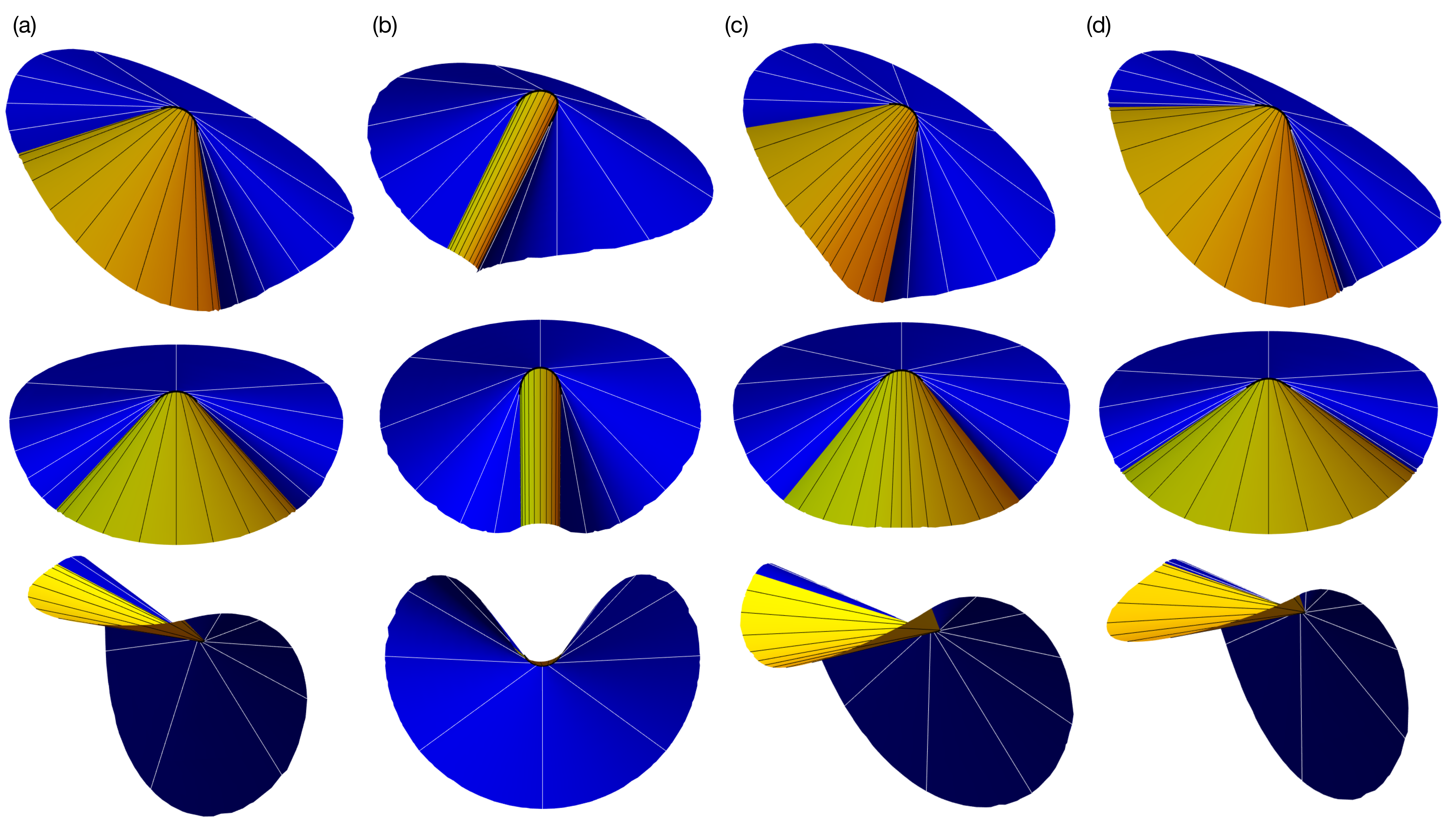}
    \caption{Examples of finite curved creases. Curved bold lines indicate the crease line, yellow (light gray) and blue (dark gray) surfaces correspond to the inner and outer surfaces respectively, and thin straight lines indicate the generators. Each row corresponds to a different viewpoint of the crease, with the first row being an off-angle view, the second a view along the symmetry line of the crease, and the third a view along the generator at which the two sides of the crease meet. (a) Crease with zero mean curvature at termination points, with $\kappa_g=1$, $\theta_0=\pi/2$, $s_t = \pi/4$, and $\phi = 3\pi/8$. (b) Crease with a cylinder as the inner surface, with $\kappa_g=1$ and $L=1$. (c) Crease with mean curvature jump at termination points, with $\kappa_g=1$, $\theta_0=\pi/2$, $s_t = \pi/4$, and $\phi = 3\pi/8$. (d) Crease with a vanishing geodesic curvature at termination points, with $\kappa_g(s)=-(s^2-s_t^2)$, $\theta_0=\pi/2$, $s_t = \pi/4$, and $\phi = 3\pi/8$.}
    \label{fig:symmcreases}
\end{figure*}

We now demonstrate a range of realizable surfaces, using several strategies. We consider specific examples of finite symmetric creases, which obey the conditions outlined in Section \ref{finitecreases}. The first example we consider is a crease formed from a circular arc, i.e. a crease with constant geodesic curvature. As explained in Section \ref{isometricity}, the crease shape will be fully determined if one specifies $\kappa_g$, $\tau$, and $\theta$ everywhere along the crease. We choose the simplest functional forms for $\theta$ and $\tau$ such that they obey all of the conditions outlined previously:
\begin{gather}
    \theta(s) = (\theta_0-\pi)\bar s^2 (\bar s^2 - 2) + \theta_0 \\
    \tau(s) = C \theta'(s) = \frac{4\phi}{s_t} \bar s (\bar s - 1)(\bar s +1)
\end{gather}
where $\bar s \equiv s/s_t$ and $\phi \equiv C(\theta_0-\pi)$ is the total twist of the crease line from the center to the tip. We can find constraints on $\phi,\, \theta_0$, and $s_t$ from the positive splay condition. At $s_t$, the result follows from Eq. \eqref{eq:nondimsplay}, and puts limits on the value of $\phi$: $(\pi-\theta_0)/2<\phi<3(\pi-\theta_0)/4$. Further constraints can be found by applying the positive splay condition at the symmetry point $s=0$, as discussed in Section \ref{finitesymmcrease}. Furthermore, we have found that if the positive splay condition is satisfied both at the symmetry point and at the termination points, it will be satisfied everywhere along the crease. An example of this type of crease is shown in Fig. \ref{fig:symmcreases}a.

\subsection{Crease around a specified surface: cylinder}
\label{cylinder}

As explained earlier in Section \ref{isometricity}, the crease shape can also be determined by specifying one surface and the geodesic curvature of the crease. Here we treat the simple case where the inner surface lies on a cylinder. The cylinder case is of particular interest, since it only marginally satisfies the positive splay condition. This surface is obtained by wrapping a flat sheet, and so is developable by construction. We again choose our crease curve to have a constant geodesic curvature, and so the curve's shape is fully determined by $\kappa_g$ and the radius of the cylinder $L$. We take the axis of our cylinder to be along the $z$-direction, and parametrize our curve as follows:
\begin{equation}
    \vec{X}(s) = \left(L \cos\left(\frac{\sin \kappa_g s}{\kappa_g L}\right), L \sin\left(\frac{\sin \kappa_g s}{\kappa_g L}\right), (1/\kappa_g) \cos\kappa_g s\right).
\end{equation}
One readily verifies that $\vec{X}'(s)$ is of unit length and is tangent to the surface of the cylinder at $\vec{X}(s)$. Furthermore the curvature vector $\vec{X}''(s)$ has a component in the tangent plane of the cylinder at $\vec{X}(s)$ of magnitude $\kappa_g$. In order for our curve to satisfy the condition $\gamma_\pm\rightarrow0$ as $s \rightarrow s_t$, the crease line's tangent $\vec{X}'(s)$ must point along the cylinder axis at $\pm s_t$. We must therefore have $\kappa_g s_t = \pi/2$ so that the curve spans a half-arc of a circle. Note that the generators on the inner surface are parallel by construction, so that they automatically satisfy the positive splay condition. The outer side of the crease will be determined as outlined in Section \ref{isometricity}, so that we can specify our crease for any given $\kappa_g$ and $L$. An example of such a crease is shown in Fig. \ref{fig:symmcreases}b.

We note that at the termination points, our crease does not satisfy the zero mean curvature condition, due to the fact that the cylindrical side has a constant non-zero curvature everywhere. As argued in Section \ref{addconstraints}, this means that the mean curvature must be discontinuous at the tip. Such discontinuities are common in physical sheets, as when a sheet is peeled away from a flat, adhesive substrate \cite{Adhesive}. This discontinuity does not prevent the construction of continuous, isostatic deformation around the crease. It leads us to study sheets with such discontinuities more generally. Therefore, we may consider other examples where there is a jump in the mean curvature at the termination points.

\subsection{Constant geodesic curvature crease with mean curvature discontinuity}
\label{discontH}

If we relax the zero mean curvature condition at the termination points, the continuous mean curvature conditions given by Eqs. \eqref{eq:thetaprimest} and \eqref{eq:taust} are no longer required. Thus we no longer need $\tau$ and $\theta'$ to vanish at $s_t$. We can then choose simpler expressions for $\theta$ and $\tau$ such that they only vanish at the symmetry point:
\begin{gather}
    \theta(s) = (\pi-\theta_0)\bar s^2 + \theta_0 \\
    \tau(s) = C \theta'(s) = -\frac{2\phi}{s_t} \bar s
\end{gather}
where again $\bar s \equiv s/s_t$ and $\phi\equiv C(\theta_0-\pi)$ is the total twist of the crease line from the center to the tip. Similar to the crease described in Section \ref{constantkappag}, $\phi$ will have bounds due to the positive splay condition, this time being $(\pi-\theta_0)/2<\phi<(\pi-\theta_0)$. We show an example of this type of crease in Fig. \ref{fig:symmcreases}c.

\subsection{Crease with vanishing geodesic curvature}

So far we have only considered cases where the geodesic curvature was constant, but we also can have a varying geodesic curvature and still satisfy our conditions at the termination points. One interesting case is when $\kappa_g$ vanishes at $s_t$, so that the crease line straightens as we approach the termination points. This will affect our termination conditions, as we shall now see.

The mean curvature near $s_t$ has the same form as before, given by Eq. \eqref{eq:Hnearst}. However, we now must take into account what happens when $\kappa_g$ vanishes at $s_t$. Let us assume $\tau$ is a constant at $s_t$, and that $\theta$ and $\kappa$ have the leading order behaviors $(s-s_t) + \pi$ and $(s-s_t)^\alpha$ respectively (so $\theta'$ is constant at $s_t$ as well). From Eq. \eqref{eq:gammanearst}, we see that $\gamma_\pm(s)\sim(s-s_t)^{\alpha+1}$, so that $\gamma_\pm'(s)\sim(s-s_t)^{\alpha}$. If we substitute these scalings into Eq. \eqref{eq:Hnearst}, we find that the denominator vanishes while the numerator does not, so that the mean curvature diverges at the termination point. In order to avoid this divergence, we must have the numerator of Eq. \eqref{eq:Hnearst} vanish in such a way that $H_\pm(s_t,v)$ is constant. Therefore, both $\tau$ and $\theta'$ must vanish at $s_t$, as they did when we required $H$ to be smooth. How fast exactly they vanish is important, as we will now see.

For simplicity, we now assume that $\tau$ and $\theta$ have the same leading order behavior near $s_t$, i.e. $\tau(s)\sim\theta'(s)\sim(s-s_t)^\beta$, so that the numerator of Eq. \eqref{eq:Hnearst} has this same behavior. Assuming $\kappa_g$ near $s_t$ has the same behavior as above, we can substitute these new scalings into Eq. \eqref{eq:gammanearst} to find $\gamma_\pm(s)\sim(s-s_t)^{\alpha+1}$. Differentiating this leads to $\gamma_\pm'(s)\sim(s-s_t)^\alpha$, so that $\gamma_\pm'$ has the same scaling as $\kappa_g$. Since $\gamma_\pm(s)$ vanishes at $s_t$ faster than $\gamma_\pm'(s)$ (and therefore $\kappa_g(s)$), we can ignore it in the denominator of Eq. \eqref{eq:Hnearst}, so that the denominator goes as $(s-s_t)^\alpha$. Therefore, in order to have $H_\pm(s_t,v)$ be a nonzero constant, we must have $\alpha=\beta$. In other words, $\tau$ and $\theta'$ must vanish just as fast as $\kappa_g$ at the termination points. This will result in a jump in the mean curvature, since as mentioned before, the mean curvature of the inner and outer surfaces are of opposite sign. Furthermore, if we have $\alpha<\beta$, so that the numerator of Eq. \eqref{eq:Hnearst} vanishes faster than the denominator, then the mean curvature will vanish at $s_t$, and so will be continuous from one side of the crease to the other. This actually is a generalization of our zero mean curvature condition for the non-vanishing geodesic curvature case: $\tau$ and $\theta'$ must vanish faster than $\kappa_g$, whether $\kappa_g$ vanishes or not at $s_t$. We show an example of this particular crease with vanishing geodesic curvature in Fig. \ref{fig:symmcreases}d, using the same functional forms for $\tau$ and $\theta$ as in Section \ref{constantkappag}.

\section{Crease Energetics}
\label{energetics}

As noted above, the final shape of the creased surface can be altered by external forcing; our conditions of isometry are not sufficient to determine the shape. The actual shape with given forcing is that of minimal bending energy. In this section we investigate how the crease parameters affect this energy.

\begin{figure}
    \includegraphics[width=0.45\textwidth]{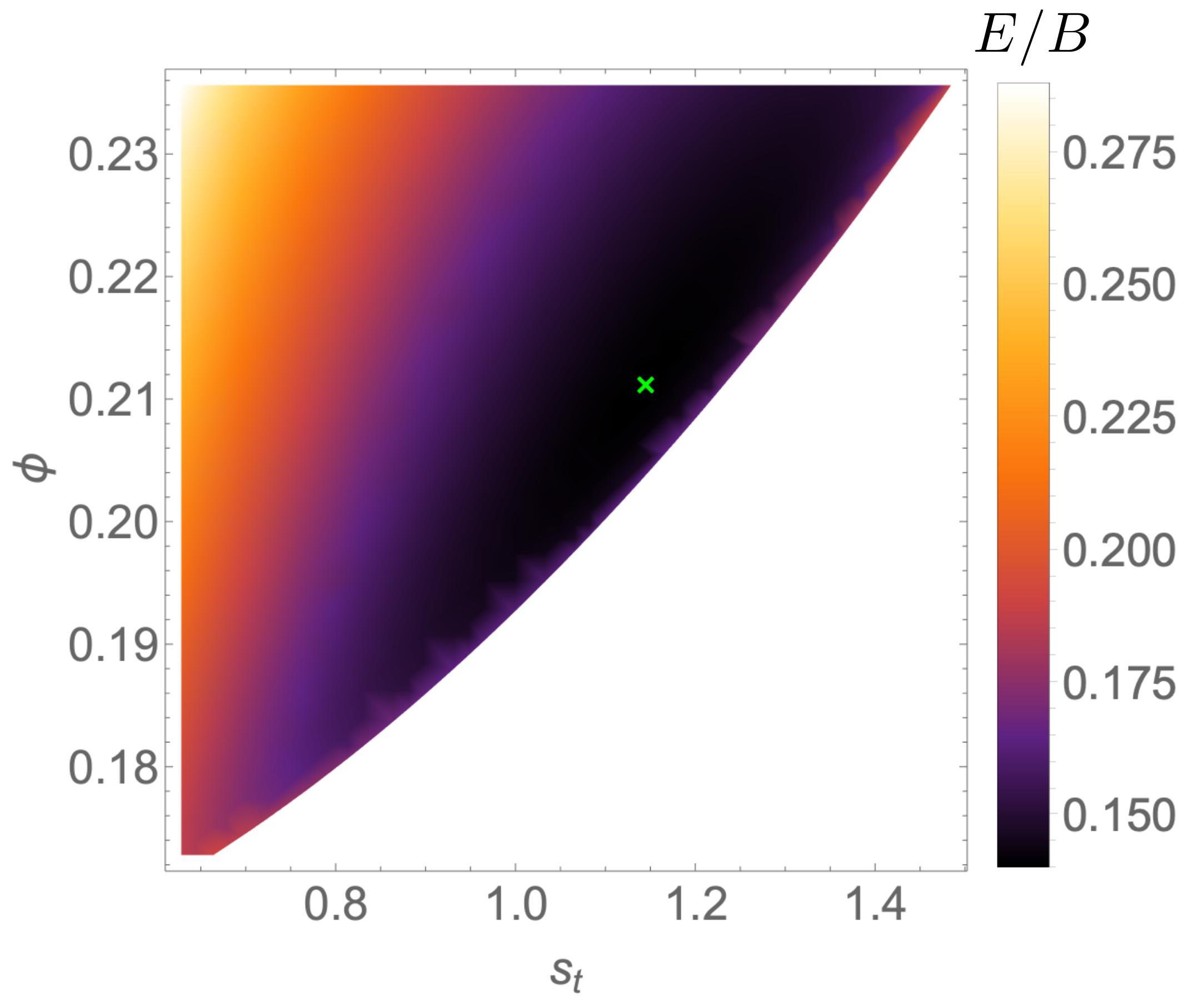}
    \caption{Heatmap of energy for a finite size crease with $\kappa_g=1$, $R=10$, and $\theta_0=0.9\pi$. Darker colors indicate a smaller energy, with the minimum indicated by a light green (light gray) cross. The blank region in the lower right hand part of the plot indicates the region where the positive splay condition on the inner surface is no longer satisfied.}
    \label{fig:HeatmapFiniteR}
\end{figure}

In order to examine the bending energy resulting from curved creases, we turn back to our first example of a constant geodesic curvature crease described in Section \ref{constantkappag}. This crease is described by three parameters: $\theta_0$, $s_t$, and $\phi$. If we compare to the d-cone, $\theta_0$ is related to the d-cone deflection, and so is a parameter we can control. Then for a given $\theta_0$, we can find values of $s_t$ and $\phi$ which minimize the total bending energy (found by integrating Eq. \eqref{eq:creaseenergy} from one end of the crease to the other), subject to the constraint imposed by the positive splay condition. An example heatmap of the bending energy is shown in Fig. \ref{fig:HeatmapFiniteR}.

\begin{figure*}
    \includegraphics[width=\textwidth]{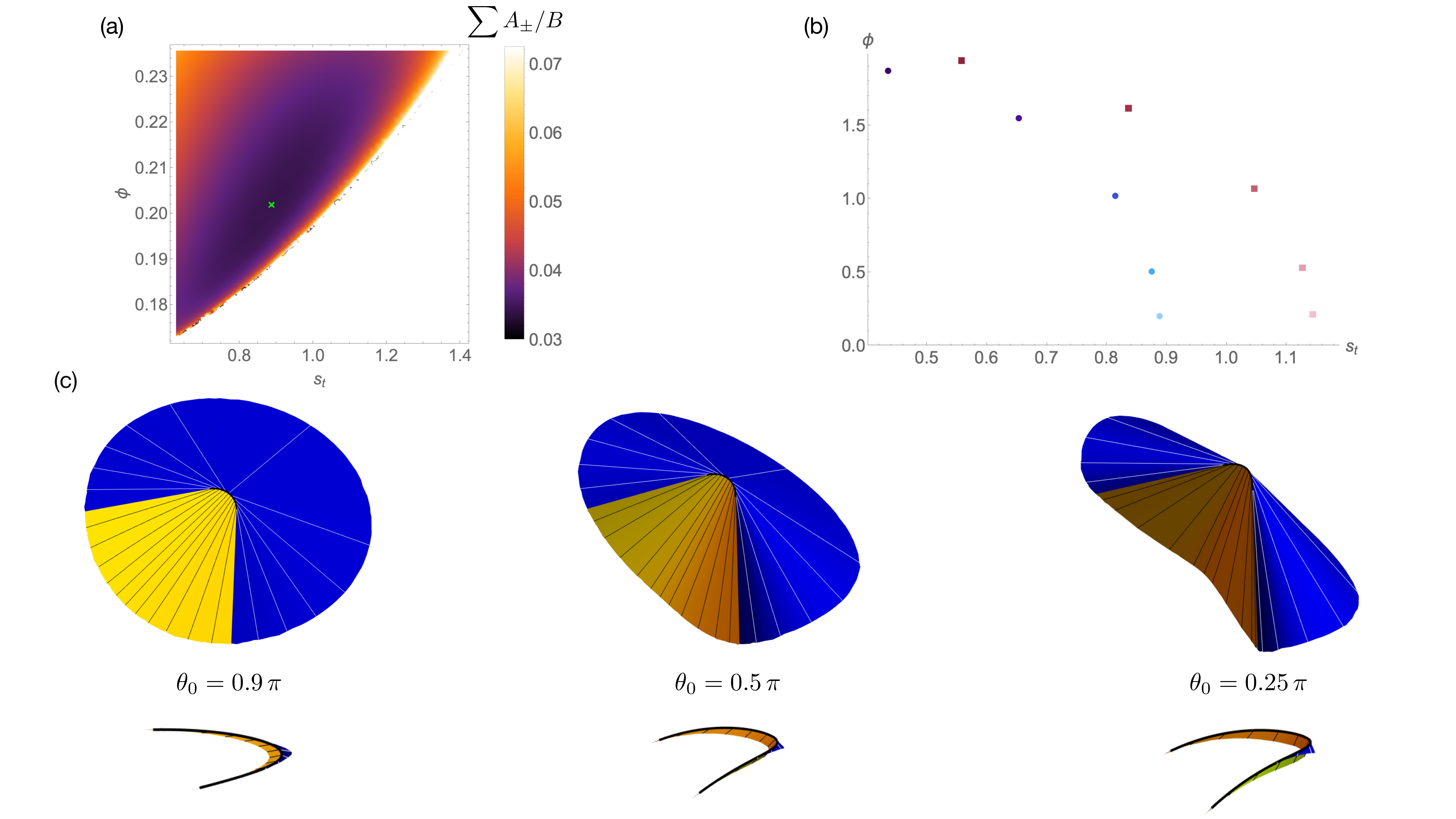}
    \caption{(a) Heatmap of $A_\pm$ for crease with $\kappa_g=1$ and $\theta_0=0.9\pi$. Darker colors indicate a smaller prefactor, with the minimum indicated by a light green (light gray) cross. The blank region in the lower right hand part of the plot indicates the region where the positive splay condition on the inner surface is no longer satisfied. (b) Parameter values which minimize the bending energy. The red squares are the values which minimize the the energy for a finite size sheet with $R=10$, while the blue circles correspond to the values which minimize $A_\pm$. Values of $\theta_0$ used are $0.1\,\pi$, $0.25\,\pi$, $0.5\,\pi$, $0.75\,\pi$, and $0.9\,\pi$, and lighter colored points correspond to larger values of $\theta_0$. (c) Equilibrium shapes of finite creases for different $\theta_0$. Bottom row shows the shape of the crease, which becomes more twisted as $\theta_0$ gets smaller.}
    \label{fig:Energetics}
\end{figure*}

However, the bending energy also depends on the size of the sheet. This means that the optimal values of $s_t$ and $\phi$ may in general depend on our chosen size of the sheet. Since we are considering asymptotically large sheets, we want to be able to minimize the energy independently of $R$. In the large $R$ limit, the bending energy is given by
\begin{equation}
    \label{eq:creaseenergylargeR}
    E \approx \frac{B}{8}\sum_\pm\mp\int\,ds\frac{(\tau\pm\theta'/2)^2 \sec^2\gamma_\pm}{\kappa_g\pm\gamma_\pm'}\log\left(\mp\frac{R(\kappa_g\pm\gamma_\pm')}{\sin\gamma_\pm}\right),
\end{equation}
which clearly has a logarithmic dependence on $R$ on each side of the crease. We can separate out this logarithmic dependence by rewriting the integrand above in the form $a_\pm(s)\cdot\log (R/d_\pm(s))$, where
\begin{equation}
    \label{eq:apm}
    a_\pm(s) \equiv \mp\frac{(\tau\pm\theta'/2)^2 \sec^2\gamma_\pm}{8(\kappa_g\pm\gamma_\pm')}
\end{equation}
and $d_\pm(s)$ is the distance to striction curve defined in Section \ref{isometricity}. We then separate the logarithm into two terms and integrate each of them separately over the extent of the crease, resulting in $E \approx B\sum_\pm A_\pm (\log R - D_\pm/A_\pm)$, where $A_\pm \equiv \int a_\pm(s)\,ds$ and $D_\pm \equiv \int a_\pm(s)\log d_\pm(s)\,ds$. Finally, defining $\log r_\pm \equiv D_\pm/A_\pm$ and combining the logarithms, we get
\begin{equation}
    \label{eq:integratedcreaseenergy}
    E \approx B\sum_\pm A_\pm \log\frac{R}{r_\pm}.
\end{equation}
Since $R$ is large, the logarithm dominates the prefactor in the expression, which has no $R$ dependence. If we remove the logarithmic dependence from Eq. \eqref{eq:integratedcreaseenergy}, we can instead minimize the sum of the prefactors $A_\pm$ in order to find the equilibrium shape of a crease in the large $R$ limit. An example heatmap of the integrated prefactor is shown in Fig. \ref{fig:Energetics}a. We can therefore find values of $s_t$ and $\phi$ that minimize $A_\pm$ extrapolated to infinite $R$ for several chosen values of $\theta_0$, as shown by the blue circles in Fig. \ref{fig:Energetics}b, and see that as the crease angle becomes more pronounced, the preferred crease length becomes shorter and the crease twists more out of plane (this is visually more obvious in the shapes shown in Fig. \ref{fig:Energetics}c). This increasing twist is also observed in other known creases, such as the narrow circular creases in \cite{DiasPRL}, where the crease buckles out of plane as one folds along the mid-line. These trends are also observed when minimizing the energy of finite-size sheets, as shown by the red squares in Fig. \ref{fig:Energetics}b.

We can also compare the energies of the different examples constructed in Section \ref{examples}. For this comparison, we use the shapes shown in Fig. \ref{fig:symmcreases}, whose parameters were chosen so that they had similar opening angles, crease lengths, and twist. We find that the energies of the shapes a, c, d are nearly the same, with the energy of the outer surface being around an order of magnitude larger than the inner energy. The remaining shape, whose inner surface is a cylinder, however exhibits a very different behavior. While we chose a cylinder radius so that the opening angle and twist angle were comparable to those of the other examples, the crease length will be fixed for a given $\kappa_g$, as explained in Section \ref{cylinder}. Furthermore, because the inner surface is a cylinder, its bending energy will not depend logarithmically on $R$, but instead linearly, so that the inner energy dominates for large sheets.

We have said in Section \ref{crescentcrease} that a finite curved crease can potentially describe the crescent observed in the d-cone core, but we have not yet justified why this proves to be better than previous approaches. In order to do so, we will compare it to the traditional picture of the d-cone. As described in Section \ref{dcone}, the traditional picture treats the outer and core regions separately, where the outside is an isometric cone whose generators have a common intersection point, and the core is a region of stress focusing whose size $R_c$ is set by a balance between bending and stretching. The total (bending) energy of the outer cone region is given by Eq. \eqref{eq:dconeEB}, and exhibits a logarithmic dependence on the sheet size $R$. As we noted above, in the large $R$ limit, the bending energy of a curved crease has a logarithmic dependence on $R$ as well, in the form $E\approx B\sum_\pm \int ds\,a_\pm(s)\cdot\log (R/d_\pm(s))$. If we compare this expression to Eq. \eqref{eq:dconeEB}, we see that at each point $s$ along the crease there is an effective $R_c$, which is the distance from the crease to the striction curve, $d_\pm(s)$, and is non-zero everywhere along the crease except at the two termination points. The vanishing of $d_\pm(s)$ at the termination points does lead to divergences of the energy density at these points, but these divergences are in general integrable, so that the total energy of the crease will be finite. Additionally, as we showed above, one can fully separate out the $R$ dependence in the energy, leading to Eq. \eqref{eq:integratedcreaseenergy}, which shows that each side of the crease has a global effective $R_c$, given by $r_\pm$. In the traditional approach, one excludes the d-cone core region when calculating the bending energy of the overall structure to avoid a divergence at the tip. However, for the crease, in the zero-thickness limit, its bending energy will be finite without having to exclude any region around the crease line. This suggests that viewing the d-cone as a finite curved crease is more advantageous than previous approaches, as it can describe the d-cone core crescent along with the entire surrounding material with just its bending energy, without imposing a cutoff near the crescent.

\section{Discussion}
\label{discussion}

We have outlined conditions necessary to have a finite curved crease in an infinite unstretchable sheet. We showed that the shape of the crease dictates and is dictated by the shape at infinity. The shape at infinity is essentially conical, since its generators converge to the localized crease. Accordingly it can be specified by giving a polar angle for each material azimuthal angle at infinity. For any such conical shape, the corresponding crease shape is determined.

In particular, the crease shape is determined for the d-cone shape, defined by imposing a maximum polar angle at infinity. Our methods predict the crease shape in the thin-sheet limit. We expect measurements of real d-cones to converge to the predicted shape.

The present paper stops short of providing this explicit prediction. We have also not indicated how these isometric predictions are altered by the strain effects that must appear at nonzero thickness. In particular, we have not directly addressed how our approach may help understand the observed scaling of the d-cone core radius. We have not resolved the paradox of how the core radius can depend on the outer confinement radius.

Though we have not predicted the full crease shape, certain simple predictions are readily apparent. First, as described in Section \ref{isometricity}, the binormal vector of the crease line $\hat{b}$ will be parallel to the line bisecting the opening angle of the crease. This means the normal vector of the crease line $\hat{n}$ must be perpendicular to this bisector. If one measures the shape profile of the d-cone core crescent, $\hat{n}$ can be determined everywhere along the crescent. One can then measure the opening angle profile, determine its bisector along the crease, and compare it to $\hat{n}$.

A second test involves the curvatures of the surfaces on either side. One can measure the mean curvature profile of the surface, particularly at the symmetry line of the crescent, where the generators will be orthogonal to the crescent. At the crescent itself, the mean curvature will be just half the normal curvature $\kappa_N$, per Eq. \eqref{eq:meancurv}. Then combined with the opening angle $\theta$ and the crescent curvature $\kappa$ at the symmetry point, one can verify whether the compatibility condition of Eq. (\ref{eq:geod}b), i.e. $\kappa_{N\pm} = \pm \kappa \cos \theta/2$, holds on both sides. We note especially that the inner and outer curvatures are equal in magnitude at the crease at this symmetry point. A similar analysis could also be done at any point in the surface by comparing to the expected mean curvature given by Eq. \eqref{eq:meancurv}, though one would then have to know the generators in the surface.

A third observation involves the tip of the crescent. Our analysis shows that the mean curvature of a crease with a non-negative $\kappa_g$ cannot change sign except at the termination point, and the generators must be parallel to the tangent vector there. This implies that the crescent line points along the inflection line of a d-cone. Finally, the curvature $\kappa(s)$ and torsion $\tau(s)$ of the crescent can be determined from the measured crescent profile, as well the opening angle $\theta(s)$. Our construction would then be able to infer the outer shape and compare it with the expected circular outer shape of a real d-cone.

We have so far only talked about the energy cost due to bending of the sheet around the crease. If there were no external forces involved, such as from folding along the crease line or confining the sheet, the equilibrium shape of the material would just be a flat sheet. When the crease forms, the surrounding material bends and acquires elastic energy, but the formation of the crease itself in the material must cost some amount of energy as well. One can account for this, for instance, by assigning the crease a stiffness and an equilibrium angle at which the crease's energy cost is minimized, as done by \cite{DiasPRL}.

Additionally, in real sheets the thickness is no longer zero, and there will be stretching in addition to bending, which will be localized around the crease. There are two known cases which account for stretching near a line singularity in a thin sheet. The first is the stretching ridge described in Section \ref{crumpling}. However, a stretching ridge is always straight, and so cannot account for the curvature of the crease line. The other case is known as the ring ridge, which appears when one deforms a spherical shell until a dent forms \cite{Pogorelov}. In the formation of this ridge, the ring will radially move inward, resulting in a compressive strain along the circle that is linear in the ridge width. This differs from the stretching ridge, where the strain is due to a transverse displacement of the ridge line, and so is quadratic in the ridge width. While the curved crease does not exactly map to the ring ridge, in the nearly-flat case, a curved crease will be a finite arc of a circle, in which case a similar analysis can be applied.

If we consider a curved crease as a segment of a ring ridge whose width is $w$, the induced strain $\gamma$ will be proportional to $\kappa_g w$. Since the area is proportional to $s_t w$, the stretching energy is of order $E_s \sim (B/h^2) \gamma^2 s_t w \sim B \kappa_g^2 w^3 s_t/h^2$. There is also a bending energy due to the ridge width, which is of order $E_b \sim B w^{-2} s_t w \sim B s_t/w$. Extending this analysis to a non-flat crease is not straightforward, since as one decreases the opening angle, the crease line's curvature is no longer constant and its torsion is no longer zero. This is demonstrated in Fig. \ref{fig:Energetics}b, as we see that the twist of the crease is comparable to the angular extent of the crease. Therefore, we do not treat the general case here. Furthermore, it is not obvious how these energies can explain the observed scaling of $R_c$, based on the reasons given earlier in this section.

In Section \ref{energetics}, we discussed the subtlety of minimizing the bending energy in the asymptotically large $R$ limit, as the equilibrium shape of a crease will in general depend on the sheet size. It would seem the same should be true for a d-cone. Experiments and simulations are limited in the sheet thickness-to-size ratios they are able to probe, and so cannot explore the asymptotically large $R$ limit. While others have consistently observed the core size scaling $R_c \sim h^{1/3}R^{2/3}$, this may not be the true scaling for asymptotically large sheets. Therefore, it is possible that if previous studies were carried out with a much larger dynamic of thickness-to-size ratios, one would observe a core size is proportional to the thickness, which would be consistent with energy balance arguments.

Finally, we would like to make some comparisons between our curved crease structures and d-cones. We have shown several examples of terminating creases in Fig. \ref{fig:symmcreases}, which all have crease lines similar to the observed core crescent. However, the outer behavior is clearly different: the outer surfaces are not conical, but instead flatten out towards the symmetry line of the crease. This is to be expected, since we do not implement any constraints to support the other side like in a d-cone, but instead constrain the shape of the crease line. One could enforce the outer surface to be a cone geometrically, but one must be careful about the liftoff region, as this occurs on the outer surface, so the entire outer surface of the crease cannot be a cone. One could also adjust the functional forms for $\kappa_g$, $\tau$, and $\theta$ in order to adjust the curvature of outer surface while still satisfying the termination conditions. However, we are still forced to acknowledge that the crease picture is incomplete, in terms of determining the length of the crease line. While we are able to find equilibrium shapes of curved creases, the finite thickness of the sheet is never considered, and so we cannot expect to learn about the dependence of the crescent length on the thickness without something additional.

\section{Conclusion}

We believe our approach of treating the d-cone core crescent as a curved crease may give new insight into understanding the energetics of the core region. One does not have to separate the core region from the outer conical region as previous approaches do. This seems more promising to understanding why the core radius could have any dependence on the outer dimensions of the sheet. Furthermore, the core crescent serves as an example of a singular region of stress-focusing whose geometry in the zero-thickness limit is not known a priori. Our approach provides a way forward to determine this geometry, and subsequently to understand the effects of stretching in the crescent in real finite-thickness sheets. There are also other similar structures that may be studied by our approach. An example is the Pogorelov ring ridge: when one deforms the dimple more and more, at some point it is energetically favorable for the ring to buckle, forming a polygonal ridge with straight ridges meeting at sharp corners. These corners have similarities to the d-cone core crescent. Another example are the forced vertices studied by Gottesman et al \cite{GottesmanNatComm}. These are formed in a way similar to that of a d-cone, but may exhibit different core radii based on the forcing protocol. By examining the exact geometry of the crescent via curved creases, we hope the observed scaling of $R_c$ can be properly reconciled.

\begin{acknowledgments}

The author would like to thank Marcelo Dias, Enrique Cerda, Shankar Venkataramani, and Anshuman Pal for illuminating discussions. The author also thanks Thomas A. Witten for his guidance, support, and enlightening conversations. This work was performed as part of the author’s Ph.D. research under the supervision of Thomas A. Witten, and was primarily supported by the University of Chicago Materials Research Science and Engineering Center, which is funded by National Science Foundation under award number DMR-2011854.

\end{acknowledgments}

\appendix

\section{Scaling of mean curvature for developable surfaces}
\label{meancurvscaling}

We present here a proof that the mean curvature $H$ of a developable surface is inversely proportional to the distance along a surface generator from the striction curve. For a given surface, the mean curvature is
\begin{equation}
    \label{eq:H}
    H = \frac{lG - mF + nE}{2(EG-F^2)},
\end{equation}
where $E,F,G$ are the coefficients of the first fundamental form and $l,m,n$ are the coefficients of the second fundamental form \cite{doCarmo}. We can choose a different base curve $\vec{Y}$ of our ruled surface so that its tangent is perpendicular to $\hat{g}_\pm$, and therefore parallel to $\hat{g}_\pm'$. This follows from the fact that $\vec{Y}'$, $\hat{g}_\pm$, and $\hat{g}_\pm'$ are coplanar for a developable surface, and that $\hat{g}_\pm$ and $\hat{g}_\pm'$ are orthogonal (in particular, it follows from Eq. \eqref{eq:dpm} that $\hat{g}_\pm'=\vec{Y}'/d_\pm$). Therefore, both fundamental forms will be diagonal, so that $F=m=0$. Furthermore, since $\hat{g}_\pm$ is a direction of zero curvature, we also have $n=0$, so that the mean curvature simplifies to $H = l/2E$. We can calculate these remaining coefficients from the surface tangent vector $\vec{t}_{s\pm}$ along $s$ (using the parametrization $\vec{X}_\pm(s,v)\equiv\vec{Y}(s) + v\hat{g}_\pm(s)$) and its derivative with respect to $s$:
\begin{gather}
    \vec{t}_{s\pm} = \partial_s\vec{X}_\pm = \vec{Y}' + v\hat{g}_\pm' = (d_\pm + v)\hat{g}_\pm'\\
    \partial_s\vec{t}_{s\pm} = d_\pm' \hat{g}_\pm' + (d_\pm + v)\hat{g}_\pm''.
\end{gather}
The remaining coefficients are then
\begin{gather}
    E = \|\vec{t}_{s\pm}\|^2 = (d_\pm + v)^2\|\hat{g}_\pm'\|^2 \\ \label{eq:l}
    l = \partial_s\vec{t}_{s\pm}\cdot\hat{N}_\pm = (d_\pm + v)\hat{g}_\pm''\cdot\hat{N}_\pm,
\end{gather}
where in Eq. \eqref{eq:l} we use the fact that $\hat{g}_\pm'$ is tangent to the surface. We can then find the mean curvature using Eq. \eqref{eq:H}:
\begin{equation}
    H_\pm = \frac{\hat{g}_\pm''\cdot\hat{N}_\pm/\|\hat{g}_\pm'\|^2}{d_\pm+v}.
\end{equation}
Since the numerator cannot depend on $v$, this shows that $H$ is inversely proportional to the distance from the striction curve, $d_\pm + v$, and therefore inversely proportional to the distance from the singular point along $\hat{g}_\pm$. This is consistent with the simple known case of a conical developable surface, where the striction curve is a single point, the vertex of the cone.

\bibliography{CreasePaperBibv2}

\end{document}